\documentclass[
aps,
prl,
showpacs,
amsmath,
amssymb,
preprint,
floatfix,
lengthcheck,
superscriptaddress,
groupedaddress
]{revtex4-1}

\usepackage{graphicx}
\usepackage{mathtools}
\usepackage{braket}
\usepackage{hyperref}

\begin{document} 

\title{Cavity correlated electron-nuclear dynamics from first principles} 

\author{Johannes Flick}
\email[Electronic address:\;]{flick@seas.harvard.edu}
\affiliation{John A. Paulson School of Engineering and Applied Sciences, Harvard
University, Cambridge, Massachusetts 02138, USA}
\author{Prineha Narang}
\email[Electronic address:\;]{prineha@seas.harvard.edu}
\affiliation{John A. Paulson School of Engineering and Applied Sciences, Harvard
University, Cambridge, Massachusetts 02138, USA}

\date{\today}

\begin{abstract}
The rapidly developing and converging fields of polaritonic chemistry and quantum optics necessitate a unified approach to predict strongly-correlated light-matter interactions with atomic-scale resolution. Combining concepts from both fields presents an opportunity to create a predictive theoretical and computational approach to describe cavity correlated electron-nuclear dynamics from first principles. Towards this overarching goal, we introduce a general time-dependent density-functional theory to study correlated electron, nuclear and photon interactions on the same quantized footing. In our work we demonstrate the arising one-to-one correspondence in quantum-electrodynamical density-functional theory, introduce Kohn-Sham systems, and discuss possible routes for approximations to the emerging exchange-correlation potentials. We complement our theoretical formulation with the first \emph{ab initio} calculation of a correlated electron-nuclear-photon system. From the time-dependent dipole moment of a CO$_2$ molecule in an optical cavity, we construct the infrared spectra and time-dependent quantum-electrodynamical observables such as the electric displacement field, Rabi splitting between the upper and lower polaritonic branches and cavity-modulated molecular motion. This cavity-modulated molecular motion has the potential to alter and open new chemical reaction pathways as well as create new hybrid states of light and matter. Our work opens an important new avenue in introducing \emph{ab initio} methods to the nascent field of collective strong vibrational light-matter interactions. 
\end{abstract}

\date{\today}

\maketitle


Remarkable experiments at the interface of condensed matter physics and quantum optics have sparked recent interest in understanding strongly correlated electronic, nuclear, and electromagnetic field degrees of freedom induced by strong light-matter coupling. Experimentally different regimes including optomechanics in picocavities~\cite{benz2016}, vibrational ultra-strong coupling for chemical systems~\cite{george2016}, strong coupling of surface plasmon polaritons and molecular vibrations~\cite{memmi2017}, and the anomalous Raman response under strong light-matter coupling~\cite{shalabney2015a} have been explored. Theoretically, such strong coupling has been analyzed for cavity-controlled chemistry via a polaron-decoupling~\cite{herrera2016}, vibrationally dressed polaritons~\cite{zeb2017}, for polaritonic chemistry~\cite{flick2017,feist2017}, spectroscopy~\cite{ruggenthaler2017b} or changes in the ground-state under ultra-strong coupling~\cite{martinez2017}. 

Recently, first-principles methods such as DFT and time-dependent density-functional theory (TDDFT) have been generalized to the realm of correlated electron-photon interactions. This quantum-electrodynamical density-functional theory (QEDFT)~\cite{ruggenthaler2011b,tokatly2013,ruggenthaler2014,flick2015} treats electrons and photons on the same quantized footing. As an exact reformulation of the Schr\"odinger equation, QEDFT can predict exactly correlated electron-photon dynamics in full real-space~\cite{flick2015}, linking closely with experimental observables. QEDFT has been shown to correctly capture correlated electron-photon systems~\cite{pellegrini2015,flick2017c}, but so far has not been demonstrated for problems in strong vibrational-photon coupling as observed in recent experiments~\cite{george2016, benz2016,memmi2017}.
So far, an understanding of vibrational effects in polaritonic chemistry has remained elusive. Yet, vibrational effects play a critical role in chemical reactions, for example, altering the vibrational mode by strong light-matter coupling can directly influence the reaction potentially allowing for a site-selective chemistry~\cite{thomas2016}. Since in strong vibrational-photon coupling experiments, the vibrational energies are on the same order of magnitude as the cavity mode, theory requires treating both on the same level of theory~\cite{flick2017b}. To computationally capture the correlated nature of the electron-nuclear interaction, many different approaches have been pursued in a DFT framework ~\cite{capitani1982, kreibich2001,kreibich2008, chakraborty2008,sirjoosingh2011,baroni2001, requist2016, giustino2017}. However, none of these methods include quantized electromagnetic fields which are essential for cavity correlated effects~\footnote{Other possible approaches include exact factorization~\cite{abedi2010, min2017, hoffmann2018}, path integrals~\cite{markland2018}, and perturbative theory~\cite{baroni2001,giustino2017}.}.

We close this critical gap and present a comprehensive theory that is capable of treating electron-nuclear-photon systems on the same quantized footing. In this paper, we discuss an important generalization of QEDFT to the realm of nuclear interactions with strong implications for experiments in cavity-driven molecule-light interactions. In the nonrelativistic limit and dipole approximation, QEDFT exploits the one-to-one correspondence between internal variables, i.e. the time-dependent electron density $n(\textbf{r},t)$, and the mode-resolved electric displacement coordinate $q_\alpha(t)$ to external variables, i.e. the time-dependent external potential $v_\text{ext}(\textbf{r},t)$ and a time-dependent current $j^{(\alpha)}_\text{ext}(t)$ for given initial state.

The paper is organized as follows: we will first discuss the general setup, define the internal and external variables for a density-functional functional theory, discuss the one-to-one correspondence, and setup the Kohn-Sham system as an efficient computational scheme. As example of the predictive power of the theory, we study the case of CO$_2$ in an optical cavity that gives rise to Rabi-splitting, which we \emph{quantitatively} capture. 

\begin{figure}[ht] 
\centerline{\includegraphics[width=0.5\textwidth]{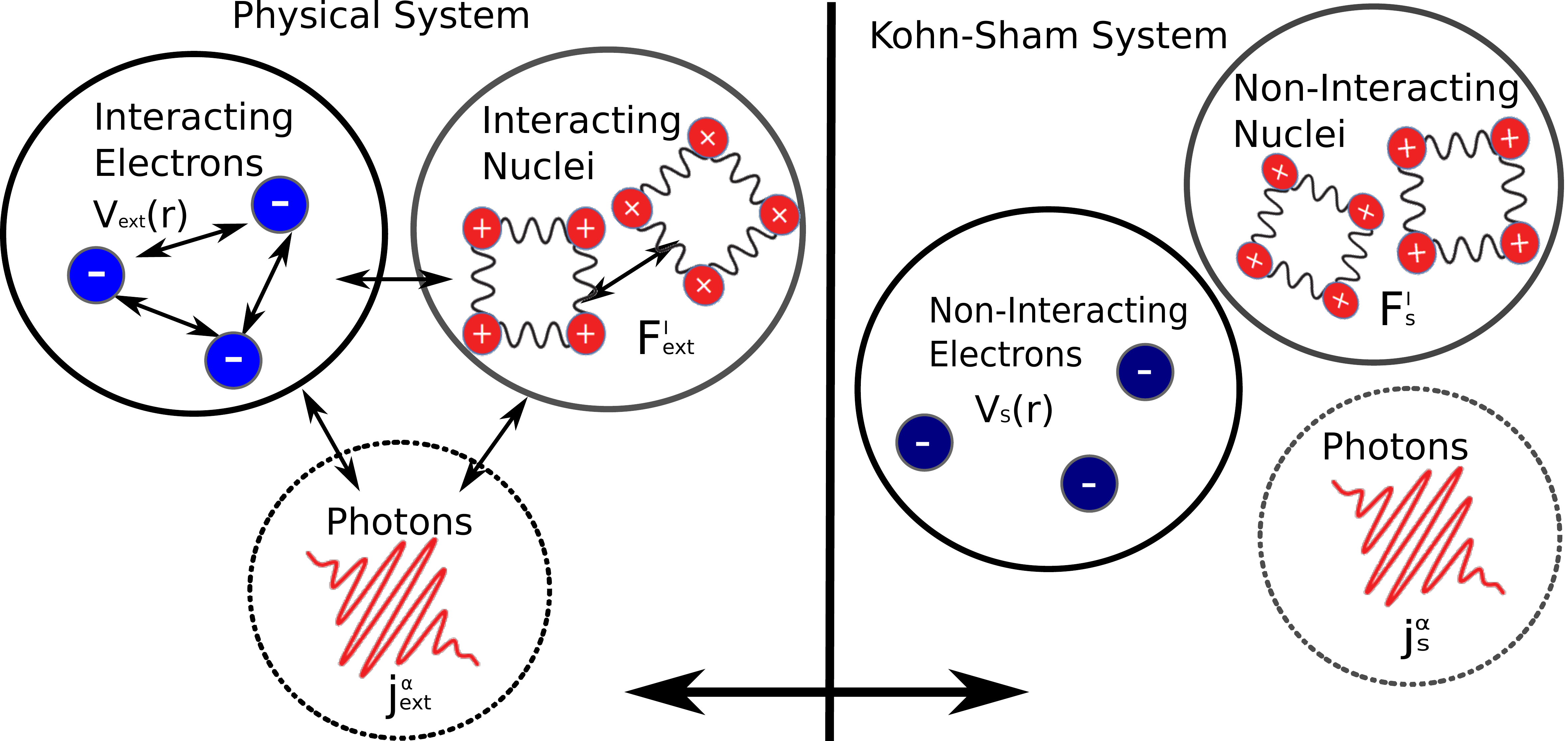}}
 \caption{\textbf{Schematic of the theoretical formalism.} The physical system consists of interacting electrons, interacting nuclei and photons controlled by the external variables, $v_\text{ext}(\textbf{r},t), F^{(I)}_\text{ext}(t)$, and $j^{(\alpha)}_\text{ext}(t)$. The physical system can be simulated by a numerically efficient Kohn-Sham system consisting of non-interacting particles with effective potentials $v_\text{s}(\textbf{r},t), F^{(I)}_\text{s}$, and $j^{(\alpha)}_\text{s}$.}
 \label{fig:00}
 \end{figure}

The general setup of the theory is as follows. The matter component of the correlated system contains $n_e$ electrons and $N_N=\sum_{I=1}^K N_I$ nuclei. With $K$ we specify the number of different nuclei species, each containing $N_I$ nuclei. We define a nuclei species $I$ by common charge $Z_I$ and mass $M_I$. If a nuclear species contains more than one nucleus, these particles are physically indistinguishable, as is the case for more than one electron. The matter component of the system is coupled to $\mathcal{N}$ quantized electromagnetic field (photon) modes. In the nonrelativistic limit, length-gauge, and dipole approximation~\cite{craig1998}, the dynamics of the system is given by the following time-dependent Schr\"odinger equation with initial state $\Psi_0$ and many-body Hamiltonian $\hat{H}(t)$~\cite{flick2017b}~\footnote{We use atomic units throughout the paper.}
\begin{align}
\label{eq:schrod}
i  \frac{\partial}{\partial t}\Psi(\underline{\textbf{r}},\underline{\textbf{R}},\underline{q},t) &= \hat{H}(t) \Psi(\underline{\textbf{r}},\underline{\textbf{R}},\underline{q},t),\\
\Psi(\underline{\textbf{r}},\underline{\textbf{R}},\underline{q},t=t_0) &= \Psi_0(\underline{\textbf{r}},\underline{\textbf{R}},\underline{q})\nonumber,
\end{align}
where we introduce the following notation for the electronic coordinates $\underline{\textbf{r}}= (\textbf{r}_1,...,\textbf{r}_{n_e})\nonumber$, the nuclear coordinates $\underline{\textbf{R}}= (\textbf{R}_{1,1},...,\textbf{R}_{K,N_K})\nonumber$, and the photon coordinates $\underline{q}= (q_1,...,q_\mathcal{N})\nonumber$, respectively~\footnote{We omit the electron and nuclear spin-index for clarity.}.

The Hamiltonian of the full problem is given by
\begin{align}
\label{eq:h-ephpt}
\hat{H}(t) = \hat{H}_0 + \hat{H}_\text{ext}(t),
\end{align}
where $\hat{H}_0$ describes the internal Hamiltonian of the different subsystems and their interactions, and $\hat{H}_\text{ext}(t)$ allows to control the entire system using external classical variables. Let us first specify
\begin{align}
\hat{H}_\text{ext}(t) &=  \int d\textbf{r} \, v_\text{ext}(\textbf{r},t)\hat{n}(\textbf{r}) + \sum_{I=1}^{K}\textbf{F}_\text{ext}^{(I)}(t) \cdot \textbf{R}_I \nonumber\\
&+ \sum_{\alpha=1}^{\mathcal{N}} \frac{j_\text{ext}^{(\alpha)}(t)}{\omega_\alpha} \hat{q}_\alpha.\label{eq:hext}
\end{align}
Hereby we have defined the external potential $v_\text{ext}$ that couples to the electron density 
\begin{align}
n(\textbf{r},t) = \biggl\langle{\Psi(t)}\biggl|\sum_{i=1}^{n_e}\delta(\textbf{r}-\textbf{r}_i)\biggl|{\Psi(t)}\biggl\rangle.
\end{align}
where the many-body wave function $\Psi(t)$ is the solution to Eq.~\ref{eq:schrod}. The classical force $\textbf{F}_\text{ext}^{(I)}(t)$ couples to
\begin{align}
\textbf{R}_I(t)=\biggl\langle{\Psi(t)}\biggl|\sum_{\beta=1}^{N_I}\textbf{R}_{I,\beta}\biggl|{\Psi(t)}\biggl\rangle.
\end{align}
For every species in the system, $\textbf{R}_I$ corresponds to the center-of-mass motion of that species. If the species contains more than a single nucleus, we find a system of indistinguishable particles and therefore the individual $\textbf{R}_{I,\beta}$ can not be told apart and only the center of mass motion is measurable~\cite{gross1996}.
Finally, the classical time-derivative of a current $j^{(\alpha)}_\text{ext}(t)$ couples to the photon displacement coordinate 
\begin{align}
\label{eq:photon-dis}
q_\alpha(t) = \bra{\Psi(t)}\hat{q}_\alpha\ket{\Psi(t)}.
\end{align}
This photon coordinate can be connected to the mode-resolved physical observables of the field, i.e. the electric displacement field $\hat{\textbf D}_\alpha(\textbf{x}) = {\sqrt{4\pi} \omega_\alpha {\boldsymbol{\lambda}_\alpha}(\textbf{x})}  \hat{q}_\alpha$ and is evaluated in Eq.~\ref{eq:h-pt} at the center of charge of the electron-nuclear system. The total fields follows as $\textbf{D}(\textbf{x},t) = \sum_{\alpha=1}^\mathcal{N}\langle\hat{\textbf D}_\alpha(\textbf{x}) \rangle$. For the following discussion, we assume the internal Hamiltonian $\hat{H}_0$ is of the following form
\begin{align}
\hat{H}_0 =& \sum_{i=1}^{n_e}-\frac{\vec{\nabla}_i^2}{2}+ \sum_{i>j}\frac{1}{|\textbf{r}_i-\textbf{r}_j|}+ \hat{H}_{p}\nonumber\\
+ &\sum_{I=1}^{{K}}\sum_{\beta=1}^{N_I}-\frac{\vec{\nabla}^2_{I,\beta}}{2M_I} + \hat{V}(\underline{\textbf{r}},\underline{\textbf{R}}), \label{eq:internal}
\end{align} 
where the first line describes the electronic and photonic Hamiltonian, and the second line the nuclear Hamiltonian including $\hat{V}$ that contains all electron-nuclear and nuclear-nuclear interactions~\footnote{$\hat{V}$ in Eq.~\ref{eq:internal} only include position-dependent interactions, i.e., $\hat{V}$ does not contain momentum operators. The exact form of the nuclear coordinates and the interaction in $\hat{V}$ depends on the setup of the specific system that is considered. For near-equilibrium situations with well-defined geometries, normal coordinates, i.e. phonons or vibrations, are most appropriate. In contrast, in the dissociation limit, the full \emph{ab initio} Hamiltonian is necessary to describe the problem. To capture both, we will for now not specify $\hat{V}$, but only specify it in the actual application.}. We proceed by defining the photonic Hamiltonian as
\begin{align}
\label{eq:h-pt}
\hat{H}_{p} &= \sum_{\alpha=1}^{\mathcal{N}}\frac{1}{2}\left[\hat{p}^2_{\alpha}+\omega^2_{\alpha}\left(\hat{q}_{\alpha}+\frac{\boldsymbol{\lambda}_{\alpha}}{\omega_{\alpha}} \cdot \hat{\boldsymbol \mu} \right)^2\right],
\end{align}
with the total dipole moment of the system $\hat{\boldsymbol\mu} = \sum_{I=1}^K Z_I\textbf{R}_I - \sum_{i=1}^{n_e} \textbf{r}_i$.\\
We now demonstrate that QEDFT can be extended to include nuclear systems. This generalization is based on an extension of the Runge-Gross theorem to arbitrary multicomponent systems~\cite{li1986} that has been applied to electron and nuclei coupled systems~\cite{gross1996}. We will use the arguments of Ref.~\cite{gross1996} to extend QEDFT to arbitrary correlated systems consisting of electrons, nuclei and the quantized electromagnetic field. 

Every density-functional theory is based on a one-to-one correspondence between internal variables and external variables. Both directly follow from the external Hamiltonian given by Eq.~\ref{eq:hext}. Therefore, the main formal result of this work can be illustrated by the following one-to-one correspondence that holds for a given initial state $\Psi_0$
\begin{align}
\label{eq:one-to-one}
\bigl(n,\textbf{R}_I,q_\alpha \bigl) \underset{1:1}{\longleftrightarrow} \bigl(v_\text{ext},\textbf{F}^{(I)}_\text{ext},j^{(\alpha)}_\text{ext} \bigl).
\end{align}
While the previously introduced Eqns.~\ref{eq:schrod}-\ref{eq:h-pt} define the mapping $\bigl(v_\text{ext},\textbf{F}^{(I)}_\text{ext},j_\text{ext}^{(\alpha)} \bigl) \longrightarrow \bigl(n,\textbf{R}_I,q_\alpha\bigl)$, the inverse mapping does not exist in general.

To show Eq.~\ref{eq:one-to-one}, we now introduce the equations of motion (EOM) for the internal variables in Eq.~\ref{eq:one-to-one}. We start by discussing the EOM for the photon coordinate $q_\alpha(t)$ that is given by~\footnote{As a side remark, we find identical EOM to standard QEDFT e.g. Eq.~8 in Ref.~\cite{flick2015} where we have to replace the electronic dipole moment by the total dipole moment of the system.}
\begin{align}
\label{eq:eom-photon}
\ddot{{q}}_\alpha(t) + \omega^2_\alpha  q_\alpha(t)  +\omega_\alpha {\boldsymbol \lambda_\alpha} \cdot {\boldsymbol\mu}(t)= -j^{(\alpha)}_{\text{ext}}(t)/\omega_\alpha.
\end{align}
This equation is a wave equation and identical to Maxwell's equations in the length-gauge with the external source term $-j^{(\alpha)}_{\text{ext}}(t)/\omega_\alpha$. Next, we look at the $K$ EOM for the nuclei coordinates $\textbf{R}_I$. We find
\begin{align}
M_I\ddot{\textbf{R}}_I(t) +& \sum^{N_I}_{\beta=1}\sum_{\alpha=1}^{\mathcal{N}}Z_I\omega_\alpha {\boldsymbol \lambda_\alpha} \left(q_\alpha(t)+ \frac{\boldsymbol \lambda_\alpha}{\omega_\alpha} \cdot {\boldsymbol\mu}(t)\right)\nonumber\\
+& \sum^{N_I}_{\beta=1}\textbf{F}_\text{str}^{(I,\beta)}(t)  = - \sum^{N_I}_{\beta=1}\textbf{F}_\text{ext}^{(I)}(t),\label{eq:eom-nuclear}
\end{align}
with the nuclear stress force 
\begin{align}
\textbf{F}_\text{str}^{(I,\beta)}(t) = \langle\Psi(t)|\vec{\nabla}_{I,\beta}\hat{V}(\underline{\textbf{r}},\underline{\textbf{R}})|\Psi(t)\rangle \nonumber,
\end{align}
where $\textbf{F}_\text{str}^{(I,\beta)}(t)$ is by construction identical for each particle $\beta$. 

The EOM for the electron density $n(\textbf{r},t)$ is given by the following Sturm-Liouville problem
\begin{align} 
\label{eq:eom-electron}
& \ddot{n}(\textbf{r},t) +\vec{\nabla} \cdot \textbf{F}_\text{str}(\textbf{r},t) +\sum_{\alpha=1}^\mathcal{\mathcal{N}}\vec{\nabla} \cdot \textbf{F}_\alpha(\textbf{r},t) +\vec{\nabla} \cdot \textbf{F}_N(\textbf{r},t) \nonumber\\
&= \vec{\nabla}\cdot\left(n(\textbf{r},t)\vec{\nabla} v_{\text{ext}}(\textbf{r},t)\right),
\end{align}
which contains force densities $F_{\text{str}/\alpha/N}(\textbf{r},t)$ originated by the kinetic energy, electron-electron interactions, electron-photon, electron-nuclear respectively and given by
\begin{align}
\textbf{F}_\text{str}(\textbf{r},t) =& \text{i} \bra{\Psi(t)} [\hat{T}(\underline{\textbf{r}}) + \hat{W}(\underline{\textbf{r}},\underline{\textbf{r}}'),\hat{j}_p(\textbf{r})]\ket{\Psi(t)},\nonumber\\
\textbf{F}_{N}(\textbf{r},t) =& \text{i} \bra{\Psi(t)} [\hat{V}(\underline{\textbf{r}},\underline{\textbf{R}}),\hat{j}_p(\textbf{r})]\ket{\Psi(t)},\nonumber\\
\textbf{F}_{\alpha}(\textbf{r},t)  = &  {\boldsymbol \lambda_\alpha}\bra{\Psi(t)} \hat{n}(\textbf{r}) \left(   {\boldsymbol\lambda_\alpha} \cdot \hat{\boldsymbol {\mu}}  + \omega_\alpha \hat{q}_\alpha\right)\ket{\Psi(t)}.\nonumber
\end{align}
with the paramagnetic current operator $\hat{j}_p(\textbf{r})$~\cite{ruggenthaler2011}. The kinetic energy operator $\hat{T}$, and the electron-electron interaction operator $\hat{W}$ correspond to the first and second term of Eq.~\ref{eq:internal}, respectively.\\
\indent These coupled Eqns.~\ref{eq:eom-photon}-\ref{eq:eom-electron} and the initial values $n(\textbf{r},t_0)$, $\dot{n}(\textbf{r},t_0)$, $\textbf{R}_I(t_0)$, $\dot{\textbf{R}}_I(t_0)$, $q_\alpha(t_0)$, and $\dot{q}_\alpha(t_0)$ represent an exact reformulation of the Schr\"odinger equation of Eq.~\ref{eq:schrod} and therefore completely define the internal variables of Eq.~\ref{eq:one-to-one}. The uniqueness of the mapping defined in Eq.~\ref{eq:one-to-one} can be proven under the usual TDDFT assumption of $t$-analyticity such that a Taylor expansion in $t$ around the initial time $t=t_0$ is possible. Then, we can follow closely the original TDDFT proof~\cite{runge1984} with extensions to electron-nuclear systems~\cite{gross1996} and QEDFT~\cite{tokatly2013, ruggenthaler2014}. Our proof is based on \textit{reductio ad absurdum}~\footnote{More general proofs~\cite{ruggenthaler2015b} can be formulated along the lines of e.g. the fixed-point method~\cite{ruggenthaler2011}, or the non-linear lattice Schr{\"o}dinger equation~\cite{tokatly2011}.}, thus we show that for given initial state $\Psi_0$, the assumption that there exist two different sets of external variables, i.e. $\bigl(v_\text{ext},\textbf{F}_\text{ext}^{(I)},j^{(\alpha)}_\text{ext} \bigl)$ and $\bigl(v^{'}_\text{ext},\textbf{F}_\text{ext}^{(I)'},j^{(\alpha)'}_\text{ext}) \bigl)$ that lead to the same set of internal variables leads to a contradiction. Thus, we insert
\begin{align} 
v_\text{ext}(\textbf{r},t) &= \sum_{k=0}^\infty \frac{1}{k!} v_\text{ext}^{(k)}(\textbf{r},t_0) (t-t_0)^k,\nonumber\\
\textbf{F}_\text{ext}^{(I)}(t) &= \sum_{k=0}^\infty \frac{1}{k!} \textbf{F}_\text{ext}^{(I,k)}(t_0) (t-t_0)^k,\nonumber\\
j_\text{ext}^{(\alpha)}(t) &= \sum_{k=0}^\infty \frac{1}{k!} j_\text{ext}^{(\alpha,k)}(t_0) (t-t_0)^k\nonumber
\end{align}
into Eqns.~\ref{eq:eom-photon}-\ref{eq:eom-electron} to obtain the Taylor coefficients of $(n, \textbf{R}_I, q_\alpha)$ in terms of $v_\text{ext}^{(k)}(\textbf{r},t_0)$, $\textbf{F}_\text{ext}^{(I,k)}(t_0)$, and $j_\text{ext}^{(\alpha, k)}(t_0)$ and accordingly for the second set $(n', \textbf{R}'_I, q^{'}_\alpha)$. Assuming a minimum order of $k=k_\text{min}$ for which the difference of the external set does not vanish, we find a non-vanishing difference of 
$(n, \textbf{R}_I, q_\alpha)$ and $(n', \textbf{R}'_I, q^{'}_\alpha)$ for $k_\text{min}+2$. Thus $(n, \textbf{R}_I, q_\alpha)$ and $(n', \textbf{R}'_I, q^{'}_\alpha)$ will be different at $t_0+\delta t$~\footnote{Provided the initial density $n(\textbf{r},t_0)$ is reasonably well behaved~\cite{runge1984, ruggenthaler2015b}}. Therefore two different sets of external variables $\bigl(v_\text{ext},\textbf{F}_\text{ext}^{(I)},j^{(\alpha)}_\text{ext} \bigl)$ will always lead to two different sets of internal variables, thus proving the mapping outlined in Eq.~\ref{eq:one-to-one} for given initial state $\Psi(t_0)$~\footnote{Up to a trivial phase $c(t)$ that can be added to the external potential $v_\text{ext}(\textbf{r},t)$.}. 

\indent To solve the coupled Eqns.~\ref{eq:eom-photon}-\ref{eq:eom-electron} in practice, we would need to find explicit expressions in terms of $n$, $\textbf{R}_I$, $q_\alpha$  for the nuclear force $\textbf{F}_\text{str}^{(I,\beta)}$ and the electronic force densities $\textbf{F}_\text{str}, \textbf{F}_N, \textbf{F}_\alpha$. To make approximations for the unknown forces and force densities easier, one can adopt a Kohn-Sham scheme, such that approximations in terms of the force densities of the uncoupled and noninteracting system become possible. This approach has been applied successfully to electronic-structure calculations (see, e.g., Refs.~\cite{gross1996, kohn1999}). In total, we find $n+\mathcal{N}+N_I\times K$ Kohn-Sham equations that read as follows
\begin{align}
i  \frac{\partial}{\partial t} {\varphi_i(\textbf{r},t}) &= \left[-\frac{\vec{\nabla}_i^2}{2} + v_s(\textbf{r},t) \right]{\varphi_i(\textbf{r},t})\label{eq:ks-electron} \\
M_I\ddot{\textbf{Q}}_{I,\beta}(t) & =- \textbf{F}_{s}^{(I, \beta)}(t)\label{eq:ks-nuclei}\\
\ddot{{q}}_\alpha(t)  + \omega^2_\alpha  q_\alpha(t) &=-{j^{(\alpha)}_{{s}}(t)}/{\omega_\alpha}\label{eq:ks-photon},
\end{align}
where we have to choose the same initial conditions, i.e. $n(\textbf{r},t_0) = \sum_{i=1}^{n_e} \varphi^*_i(\textbf{r},t_0)\varphi_i(\textbf{r},t_0)$, $\dot{n}(\textbf{r},t_0)$, $\textbf{R}_I(t_0) = \sum_{\beta=1}^{N_I}\textbf{Q}_{I,\beta}(t_0)$, $\dot{\textbf{R}}_I(t_0)$, and $q_\alpha(t_0)$, $\dot{q}_\alpha(t_0)$,  as in the physical system. For the photons subsystem we find the Kohn-Sham current as
\begin{align} 
\label{eq:photon-js}
{j}_{s}^{(\alpha)}(t) = \omega^2_\alpha {\boldsymbol \lambda_\alpha} \cdot {\boldsymbol \mu(t)} + {j^{(\alpha)}_\text{ext}(t)}.
\end{align} 
where all terms that are attributed to the matter-photon interaction are explicitly known, and hence the unknown expressions that take care of the proper quantum description of the matter-photon interactions are contained solely in the electronic and nuclear equations. \\
\indent In Eq.~\ref{eq:ks-nuclei}, we have introduced Kohn-Sham trajectories ${\textbf{Q}}_{I,\beta}$ for every single nucleus in the system. However, if we have indistinguishable particles, only the total trajectory ${\textbf{Q}}_{I}$ of that species is observable. In this way, the nuclear force $\textbf{F}_{s}^{(I)}(t)$ is defined such that the sum of all Kohn-Sham trajectories $\textbf{Q}_{I,\beta}$ reproduces the exact total trajectory of that species, i.e. $\textbf{R}_I(t) = \sum_{\beta=1}^{N_I}\textbf{Q}_{I,\beta}(t)$. This way we define 
\begin{align} 
\label{eq:nuclei-fs}
\textbf{F}_{s}^{(I, \beta)}(t) =
& \sum_{\alpha=1}^\mathcal{N}Z_I\omega_\alpha {\boldsymbol \lambda_\alpha} \left({q}_{\alpha}(t)+\frac{\boldsymbol{\lambda}_{\alpha}}{\omega_{\alpha}} \cdot {\boldsymbol \mu(t)} \right)\nonumber\\
&+ \textbf{F}^{(I, \beta)}_\text{Mxc}(t) + \textbf{F}^{(I)}_\text{ext}(t),
\end{align}
where the sum of $\textbf{F}^{(I,\beta)}_\text{Mxc}(t)$ is defined as $\sum_{\beta=1}^{N_I}\textbf{F}^{(I,\beta)}_\text{Mxc}(t) = \sum_{\beta=1}^{N_I}\textbf{F}^{(I, \beta)}_\text{str}(t)$ describes the exchange-correlation contribution~\footnote{If we have a electron-nuclear and nuclear-nuclear interaction $\hat{V}$ present in the system that has terms up to second order, as in the case of electron-phonon interactions with Fr{\"o}hlich coupling~\cite{giustino2017}, then we find a vanishing exchange-correlation contribution.}. For the electronic Kohn-Sham system, we define the following Kohn-Sham potentials
\begin{align}
\label{eq:qedft-vs}
v_{s}(\textbf{r},t) &= v_\text{ext}(\textbf{r},t) + v_\text{Mxc}(\textbf{r},t) 
\end{align} 
with the mean-field xc potential
\begin{align}
v_\text{Mxc}(\textbf{r},t)  = v_\text{Hxc}(\textbf{r},t) + \sum_{\alpha=1}^\mathcal{N}v^{(\alpha)}_\text{Mxc}(\textbf{r},t) + v^{(N)}_\text{Mxc}(\textbf{r},t),\nonumber
\end{align}
where these potentials are exactly defined in terms of Sturm-Liouville equations~\footnote{We notice that although we have divided the Kohn-Sham potential in Eq.~\ref{eq:qedft-vs} into contributions dominant from the electron-nuclear and electron-photon interaction, the Hartree-exchange-correlation (Hxc) potential as defined in Eq.~\ref{qedft-vhxc} also contains contributions from these interactions due to the different kinetic energy in the Kohn-Sham and the physical system.}
\begin{align}
\label{qedft-vhxc}
&\vec{\nabla}\cdot\left(n(\textbf{r},t)\vec{\nabla} v_{\text{Hxc}}(\textbf{r},t)\right) = \vec{\nabla} \cdot \left( \textbf{F}^{(s)}_\text{str}(\textbf{r},t) - \textbf{F}_\text{str}(\textbf{r},t)\right),\\
\label{qedft-vmxc-photons}
&\vec{\nabla}\cdot\left(n(\textbf{r},t)\vec{\nabla} v^{(\alpha)}_{\text{Mxc}}(\textbf{r},t)\right) = \vec{\nabla} \cdot \textbf{F}_\alpha(\textbf{r},t), \\
\label{qedft-vmxc-nuclei}
&\vec{\nabla}\cdot\left(n(\textbf{r},t)\vec{\nabla} v^{(N)}_{\text{Mxc}}(\textbf{r},t)\right) = \vec{\nabla} \cdot \textbf{F}_N(\textbf{r},t).
\end{align}

Over the last decades, the electronic-structure community has developed a large selection of possible approximations to the in general unknown exchange-correlation potential~\cite{marques2012}. In contrast, the nascent field of QEDFT has not yet seen the same development of approximations, so far only the one-photon optimized-effective potential (OEP) has been successfully derived~\cite{pellegrini2015} and applied to realistic systems~\cite{flick2017c}. Other possibilities are a parameterization along the lines of the local-density approximation (LDA)~\cite{kohn1965} in TDDFT. As being closely linked to QEDFT, the present formalism also allows to connect to the TDOEP~\cite{kuemmel2008} route that seems promising in the limit of weak and very strong electron-nuclear correlations.\\

Next, we specify the electron-nuclear potential $\hat{V}$ in Eq.~\ref{eq:internal} as~\cite{gross1996}
\begin{align}
\hat{V}(\underline{\textbf{r}},\underline{\textbf{R}}) =& \frac{1}{2}\sum_{I=1}^{K}\sum_{\beta=1}^{N_I}\sum_{J=1}^{K}\sum_{\substack{\gamma=1\\(J\gamma\neq I\beta)}}^{N_J}\frac{Z_IZ_J}{|\textbf{R}_{I,\beta} - \textbf{R}_{J,\gamma}|}\nonumber\\
& - \sum_{i=1}^{n_e}\sum_{I=1}^{K}\sum_{\beta=1}^{N_I}\frac{Z_I}{| \textbf{r}_i-\textbf{R}_{I,\beta}|},
\end{align}
where the first line describes the nuclear-nuclear interaction, while the second line describes the electron-nuclear interaction. For processes, where the overlap of nuclear wave functions remains small, such as molecular vibrations, we use the following approximation~\cite{gross1996} which can be used in Eq.~\ref{eq:ks-nuclei} 
\begin{align}
\textbf{F}_\text{M}^{(I,\beta)}(t) =& \sum_{J=1}^K\sum_{\substack{\gamma=1\\(J\gamma\neq I\beta)}}^{N_J}  \frac{Z_IZ_J ({\textbf{Q}_{J,\gamma}}- {\textbf{Q}_{I,\beta})}}{|\textbf{Q}_{I,\beta}-\textbf{Q}_{J,\gamma}|^3}\nonumber\\
 &- \int d\textbf{r} \frac{Z_I n(\textbf{r},t)\left(\textbf{r}-\textbf{Q}_{I,\beta}\right)}{|\textbf{r}-\textbf{Q}_{I,\beta}|^3}.
\label{eq:ehrenfest-force}
\end{align}
This force now depends explicitly on the individual nuclear trajectory $\textbf{Q}_{I,\beta}$ and therefore can be seen as similar as the self-interaction correction (SIC) of DFT~\cite{gross1996}. Using this equation for the matter part, we recover the Ehrenfest scheme~\cite{andrade2009}, i.e. a mixed quantum-classical scheme that treats the electrons quantum mechanically coupled to classical nuclei. Analogously, we find for the electron-nuclear potential
\begin{align}
v_\text{M}^{(N)}(\textbf{r},t) = - \sum_{I=1}^{K}\sum_{\beta=1}^{N_I}\frac{Z_I}{| \textbf{r}-\textbf{Q}_{I,\beta}(t)|}.
\end{align}

In the following, we now apply the presented formalism to vibrational strong-coupling of light to a molecular system (CO$_2$ molecule) and we present the numerical details in appendix A.
\begin{figure}[ht]
\centerline{\includegraphics[width=0.5\textwidth]{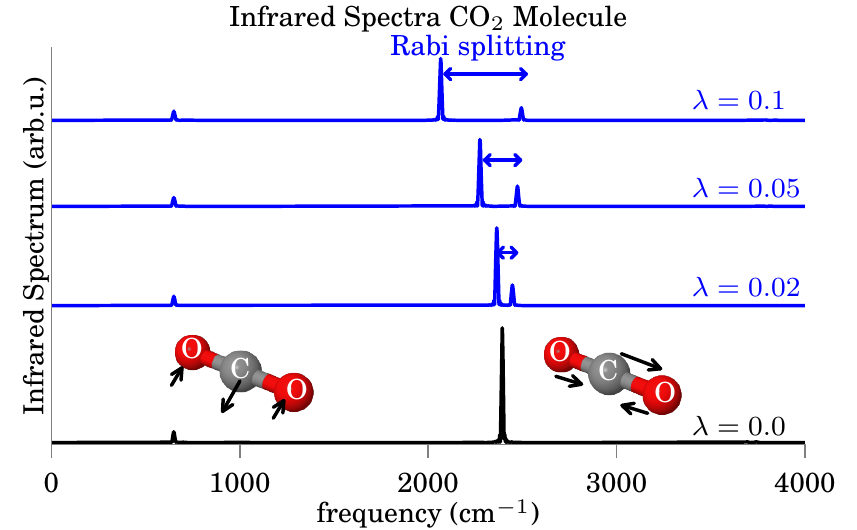}}
\caption{\textbf{Infrared spectra in vibrational strong coupling for CO$_2$.} Black spectrum refers to the spectrum outside the cavity. We explicitly depict the two infrared-active vibrational modes of the CO$_2$ molecule. Blue spectra correspond to the electron-nuclear spectrum. Importantly, we capture the Rabi splitting between the lower and upper polariton branch.}
\label{fig:01}
\end{figure}
We find for CO$_2$ three infrared(IR)-active vibrational excitations, that are shown in Fig.~\ref{fig:01} in black, one at $2430$ cm$^{-1}$ and the second one with a two-fold degeneracy at $654$ cm$^{-1}$. To obtain the infrared spectra, we initially excite the three vibrational modes such that the carbon atom is displaced by $0.01\AA$ in all three spatial directions and record the time-evolution of the total dipole moment $\boldsymbol \mu(t)$ for $5$~ps. The Fourier transform of the dipole moment yields then infrared spectrum~\cite{andrade2009}. In Fig.~\ref{fig:01}, we also depict schematically for all IR active modes their normal mode oscillation. If the molecule is now strongly coupled to a cavity mode, we find Rabi splitting in the infrared spectra emerging. To simulate vibrational strong coupling, we choose the cavity frequency $\omega_\alpha=2430 \, \text{cm}^{-1}$ to be in resonance to the vibrational excitation at $2430$ cm$^{-1}$ and with polarization in $x$-direction. By varying the matter-photon coupling parameter $ \lambda_\alpha = |\boldsymbol \lambda_\alpha |$, we can tune the system from the weak to the strong coupling limit. In Fig.~\ref{fig:01}, we show in blue the spectra for $\lambda_\alpha = (0.02,0.05,0.1)$ and we find the Rabi-splitting occurring with increasing splitting for stronger $\lambda_\alpha$.
\begin{figure}[ht]
\centerline{\includegraphics[width=0.5\textwidth]{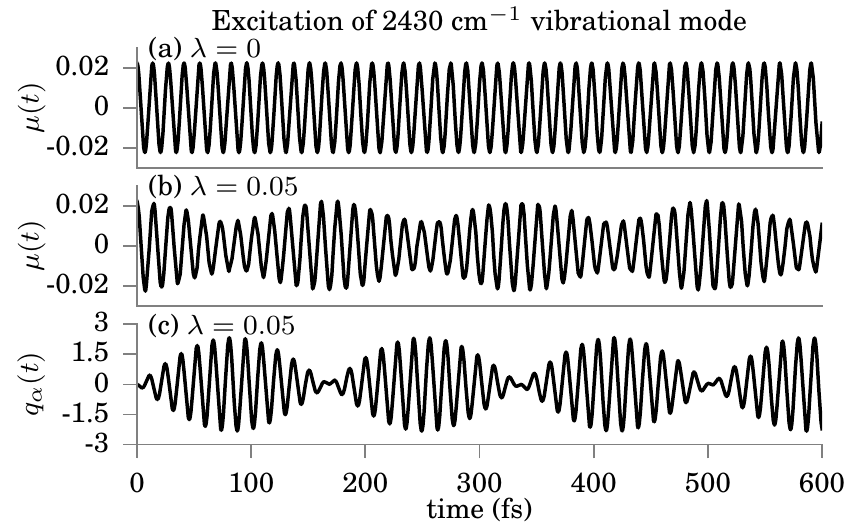}}
\caption{\textbf{Vibrational excitation at $2430$ cm$^{-1}$.} Initial displacement of the C-atom of $0.01\AA$, (a) dipole moment CO$_2$ outside the cavity, (b) dipole moment CO$_2$ under strong light-matter coupling for $\lambda_\alpha=0.05$, (c) the photon displacement coordinate $q_\alpha(t)$ as defined in Eq.~\ref{eq:photon-dis} for $\lambda_\alpha=0.05$.}
\label{fig:02}
\end{figure}
Next, to analyze the dynamics of the system under vibrational strong light-matter coupling in more detail, we initially displace the carbon molecule by $0.01\AA$ to specifically excite the $2430$ cm$^{-1}$ vibration. In Fig.~\ref{fig:02} (a), we show the time-dependent dipole moment of the system under that initial excitation for up to $600$~fs without matter-photon coupling. The system oscillates very regularly with a frequency of $2430$ cm$^{-1}$. If we now choose $\lambda_\alpha = 0.05$, we find an additional frequency occurring as an envelope that corresponds to the Rabi splitting as shown in Fig.~\ref{fig:02} (b). In (c), we show a new observable that is now possible to calculate with this novel formalism. We depict time-evolution of the photon displacement coordinate and find additionally to the regular oscillation an envelope given by the Rabi splitting.
\begin{figure}[ht]
\centerline{\includegraphics[width=0.5\textwidth]{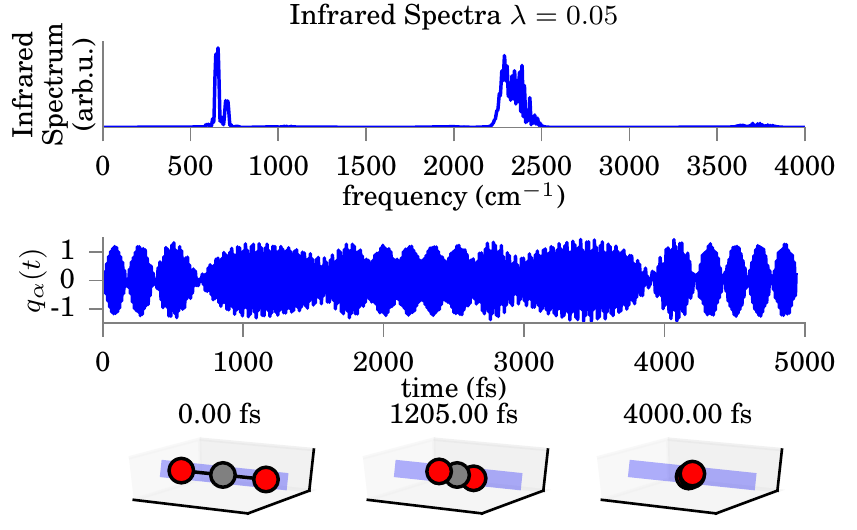}}
\caption{\textbf{Spinning molecule.} From top to bottom: Infrared spectrum after $5$~ps for a CO$_2$ molecule under strong-light matter coupling with $\lambda_\alpha=0.05$. Center: time-evolution of the expectation value of ${q}_\alpha(t)$. Bottom: Snapshots of the nuclear positions of the spinning molecule for $t=0$~fs, $t=1205$~fs, and $t= 4000$~fs. The blue area indicates the polarization direction of the photon mode.}
\label{fig:03}
\end{figure}
In the last example, we study in this paper, we choose to initialize the three nuclei with random velocities drawn from a Maxwell-Boltzmann distribution corresponding to $T=100$~K. The infrared spectrum of this run depicted in Fig.~\ref{fig:03}
shows not the same clean signature of the Rabi splitting as in Fig.~\ref{fig:01}-\ref{fig:02} but rather a broad band with many peaks around $2500$ cm$^{-1}$, although the cavity mode is in resonance to that frequency. This broad band can be understood as follows. By choosing initial random velocities, the molecule is spinning during the simulation time and thus the effective interaction strength $\lambda_{\alpha,\text{eff}}(t) = \textbf{e}_\alpha\cdot {\boldsymbol\mu}(t)$ changes strongly in time. In the center of Fig.~\ref{fig:03}, we show the expectation value of the electric displacement mode $q_\alpha(t)$. Although we also find an envelope that is a fingerprint of Rabi oscillations, in contrast to Fig.~\ref{fig:02}, we do not find a regular envelope function. This can be understood by looking at the atoms coordinates during the run as plotted in the bottom of Fig.~\ref{fig:03} for $t=(0, 1205, 4000)$~fs. Since the molecule spins around its center of mass, we find that $\lambda_{\alpha,\text{eff}}(t)\in [0,0.05]$. This directly translates into the spectra that exhibits a broadband of peaks at $2500$ cm$^{-1}$.\\

In our work we have demonstrated a new density-functional theory-based approach to treat the correlated electron-nuclear-photon problem. The Runge-Gross proof of QEDFT has been extended to the realm of nuclear motion, and we have applied this new theoretical method to analyze vibrational strong coupling, of high relevance to experimental work in this field. Our calculations are the first \emph{ab initio} calculations of vibrational strong coupling in cavities with observables that quantitatively connect with the new fields of polaritonic chemistry and nanoplasmonics that are pushing the envelope in strong light-matter interactions. Future directions include the \emph{ab initio} study of chemical reactions within the framework of polaritonic chemistry and to study excited-state phenomena~\cite{flick2018} of vibrationally strongly-coupled cavity systems. We envision using this understanding of quantum-cavity controlled vibrational strong coupling as a testbed to develop a general methodology for optical control of
chemical dynamics via strong light-matter coupling to alter the fundamental pathways of molecular species. Controlling and directing reactions in single molecule-optical cavity hybrids will provide mechanistic knowledge at the atomic-scale. These strongly-coupled molecule-cavities systems could also be used to monitor the kinetic and thermodynamic properties of chemical reactions, creating a new method of quantum correlated spectroscopy.

The authors thank Nick Rivera, and Michael Ruggenthaler for fruitful discussions. This research used resources of the National Energy Research Scientific Computing Center, a DOE Office of Science User Facility supported by the Office of Science of the U.S. Department of Energy under Contract No. DE-AC02-05CH11231, as well as resources at the Research Computing Group at Harvard University. PN acknowledges start-up funding from the Harvard John A. Paulson School of Engineering and Applied Sciences.

\bibliography{light_matter_coupling} 

\clearpage

\pagebreak
\widetext
\begin{center}
\textbf{\large Supplemental Information:\\Cavity correlated electron-nuclear dynamics from first principles}
\end{center}
\setcounter{equation}{0}
\setcounter{figure}{0}
\setcounter{table}{0}
\setcounter{page}{1}
\makeatletter
\renewcommand{\theequation}{S\arabic{equation}}
\renewcommand{\thefigure}{S\arabic{figure}}

\section{A Numerical details}
\label{sec:numerics}

To demonstrate the practicability of the presented formalism, we have implemented this scheme to the real-space TDDFT code OCTOPUS~\cite{andrade2014} is based on the combined TDDFT Ehrenfest scheme discussed in Refs.~\cite{alonso2008,andrade2009} to solve Eq.~\ref{eq:ks-nuclei} with Eq.~\ref{eq:ehrenfest-force}. To simplify the solution of the EOM for the photon mode, i.e. Eq.~\ref{eq:photon-js}, we use an analytic formula~\cite{pellegrini2015} combined with numerical integration to obtain $q_\alpha(t)$ at each time step. Additionally, we can exploit the different time scales in the system (see appendix B to speed up the simulation time that then allows us for calculations up to $5$~ps. As prototype system  we choose a single CO$_2$ molecule aligned in $x$-direction that is exposed to a single photon mode in an optical cavity. We describe the electronic structure of this molecule using the local-density approximation (LDA)~\cite{kohn1965} and treat explicitly the valence electrons of the system. The core electrons are treated implicitly using Troullier-Martins pseudopotentials~\cite{troullier1993}. We choose a spherical box of $4\AA$ in all three spatial dimensions with grid spacing of $0.1\AA$ to describe the electronic structure accurately.

\section{B Timescales in the system}
\label{sec:timescales}

For the actual numerical simulation, we exploit the different time scales present in vibrationally strongly coupled systems. If the cavity mode is in the order of an vibrational excitation, the cavity Born-Oppenheimer approximation (CBOA)~\cite{flick2017, flick2017b} can be applied. In the CBOA, we assume different timescales of the electrons and the nuclear and photons. In contrast to the latter two, the electrons can be considered fast. For a discussion of the applicability of the CBOA, we refer the reader to Ref.~\cite{flick2017b}. Exploiting these different timescales, we can use the following Lagrangian in analogy to Ref.~\cite{alonso2008,andrade2009}
\begin{align}
\mathcal{L}(\underline{\varphi},\underline{\dot{\varphi}}, \underline{\textbf{Q}}, \underline{\dot{\textbf{Q}}}, \underline{q}, \underline{\dot{q}}) &= 
\mu_e \frac{i}{2}\sum_{i=1}^{n_e}\int d\textbf{r} \varphi_i^*(\textbf{r},t) \dot{\varphi}_i(\textbf{r},t) - \dot{\varphi}_i^*(\textbf{r},t)\varphi_i^*(\textbf{r},t)\nonumber\\
&+\sum_{I=1}^K\sum_{\beta=1}^{N_I} \frac{M_I}{2}\dot{\textbf{Q}}_{I,\beta}\dot{\textbf{Q}}_{I,\beta} + \sum_{\alpha=1}^\mathcal{N} \frac{1}{2}\dot{{q}}_\alpha\dot{{q}}_\alpha - E_{KS}(\underline{\varphi},\underline{\textbf{Q}},\underline{q}),
\end{align}
where $E_{KS}$ refers to the KS energy that can be defined by the expectation value of the KS wavefunction $\Phi_s$ with the KS Hamiltonian $\hat{h}_s$ that leads to the KS equations of Eqns.~\ref{eq:ks-electron}-\ref{eq:ks-photon}. Using this Lagrangian for the nuclear and photon system, the same EOM as in Eq.~\ref{eq:ks-nuclei} and Eq.~\ref{eq:ks-photon} follow. However, the electronic Kohn-Sham equation of Eq.~\ref{eq:ks-electron} is modified to
\begin{align}
i \mu_e  \frac{\partial}{\partial t} {\varphi_i(\textbf{r},t}) =& \left[-\frac{\vec{\nabla}_i^2}{2} + v_s(\textbf{r},t) \right]{\varphi_i(\textbf{r},t})\label{eq:ks-electron-mod} 
\end{align}
where the parameter $\mu_e$, now effectively rescales the electronic velocities. If $\mu_e=1$, we recover Eq.~\ref{eq:ks-electron}. As discussed in Ref.~\cite{andrade2009}, larger values of $\mu_e$ can speed up the calculation. However, increasing $\mu_e$ increases nonadiabatic effects, as it decreases the gap of the ground-state to the excited state energy surface. For the CO$_2$ molecule, we find that values of $\mu_e=10$ are reasonable.

\end{document}